\documentclass{sig-alternate-2013}

\def\istechreport{1}

\ifdefined\istechreport
\pdfoutput=1
\fi

\usepackage[T1]{fontenc}
\usepackage[utf8]{inputenc}

\ifdefined\istechreport
\usepackage{url}
\else
\usepackage{hyperref}
\fi

\usepackage{listings}
\usepackage{graphicx}
\usepackage{subcaption}
\usepackage{enumitem}
\usepackage{amsmath}
\usepackage{array}
\usepackage{pbox}
\usepackage{todonotes}
\usepackage{xparse}
\usepackage{letltxmacro}

\LetLtxMacro{\todonote}{\todo}
\renewcommand{\todo}[2][]
{\todonote[caption={#2}, size=\footnotesize, #1]
{\renewcommand{\baselinestretch}{0.5}\selectfont#2\par}}

\setlist{itemsep=.01cm}

\renewcommand{\footnotesize}{\scriptsize}

\newfont{\mycrnotice}{ptmr8t at 7pt}
\newfont{\myconfname}{ptmri8t at 7pt}
\let\crnotice\mycrnotice%
\let\confname\myconfname%

\ifdefined\istechreport
\toappear{This work is based on an earlier work: Citizen Electronic Identities
using TPM 2.0, to appear in the Proceedings of the 4th international workshop
on Trustworthy embedded devices, \myconfname{TrustED'14}, {\mycrnotice November
3, 2014, Scottsdale, Arizona, USA, http://dx.doi.org/10.1145/2666141.2666146}}
\else
\permission{Permission to make digital or hard copies of all or part of this
work for personal or classroom use is granted without fee provided that copies
are not made or distributed for profit or commercial advantage and that copies
bear this notice and the full citation on the first page. Copyrights for
components of this work owned by others than the author(s) must be honored.
Abstracting with credit is permitted. To copy otherwise, or republish, to post
on servers or to redistribute to lists, requires prior specific permission
and/or a fee. Request permissions from permissions@acm.org.}
\conferenceinfo{TrustED'14,}{November 3, 2014, Scottsdale, Arizona, USA.\\
{\mycrnotice{Copyright is held by the owner/author(s). Publication rights
licensed to ACM.}}}
\copyrightetc{\the\acmcopyr}
\crdata{ACM 978-1-4503-3149-4/14/11 ...\$15.00.\\
http://dx.doi.org/10.1145/2666141.2666146}
\fi

\begin{document}

\title{Citizen Electronic Identities using TPM 2.0}

\ifx\isanonymous\undefined
\numberofauthors{3}
\author{
  \alignauthor Thomas Nyman\\
  \affaddr{Aalto University and}\\
  \affaddr{University of Helsinki}\\
  \email{\small{thomas.nyman@aalto.fi}}
  \alignauthor Jan-Erik Ekberg\\
  \affaddr{Trustonic}\\
  \email{\small{jan-erik.ekberg@trustonic.com}}
  \alignauthor N. Asokan\\
  \affaddr{Aalto University and}\\
  \affaddr{University of Helsinki}\\
  \email{\small{asokan@acm.org}}
}
\fi

\maketitle
\begin{abstract}
\emph{Electronic Identification} (eID) is becoming commonplace in several
European countries. eID is typically used to authenticate to government
e-services, but is also used for other services, such as public transit,
e-banking, and physical security access control. Typical eID tokens take the
form of physical smart cards, but successes in merging eID into phone operator
SIM cards show that eID tokens integrated into a personal device can offer
better usability compared to standalone tokens. At the same time, trusted
hardware that enables secure storage and isolated processing of sensitive data
have become commonplace both on PC platforms as well as mobile devices.

Some time ago, the Trusted Computing Group (TCG) released the version 2.0 of the
Trusted Platform Module (TPM) specification. We propose an eID architecture
based on the new, rich authorization model introduced in the TCGs TPM 2.0. The
goal of the design is to improve the overall security and usability compared to
traditional smart card-based solutions. We also provide, to the best our
knowledge, the first accessible description of the TPM 2.0 authorization model.
\end{abstract}

%
%
%
%

\section{Introduction}
\label{sec:introduction}

Identification and authentication of citizens using \emph{Electronic
Identification} (eID) mechanisms is beginning to be deployed in several
European countries. In some countries, such as Estonia, eID tokens are
already used widely, while eID use is taking hold in other countries
such as Belgium, Austria and Spain~\cite{Martens10}. The typical use of
eID is to access government services such as filing tax returns,
accessing healthcare records or digitally signing applications for
state benefits. However, since eID tokens provide strong identity
verification, they are also used for other services, both public and
private. For example in Estonia, more than 100 000 persons use their
eID cards as a form of public transport tickets~\cite{Martens10}. They
are also used as substitutes for driving licenses and e-banking tokens,
as well as physical access control tokens for libraries and swimming
pools~\cite{Myhr08}. Furthermore, personal computer operating systems
allow for seamless integration of PKCS\#15 cards which makes it
possible to use them with common applications such as e-mail clients
with S/MIME support.

Identity tokens, referred to as ``Electronic Signature Products''
(ESPs) by the relevant EU directives, must be trustworthy. ESPs are
logically isolated security tokens that use a signature key to
digitally sign a request on behalf of a user and thereby prove the user
identity. The ESPs are under the control of users, and can be carried
by them or be integrated into their devices. The ESP is, however, not
necessarily a physical smart card --- so far, the ESP function has also
been successfully integrated into operator SIM
cards\footnotemark\textsuperscript{,}\footnotemark, as well as in
network services controlled by password-based access. With the
increased usage of personal user devices, integrating an eID mechanism
securely also in these devices will help improve usability of such eID
compared to standalone eID tokens.

\footnotetext[1]{Estonian Mobile-ID:
\url{http://e-estonia.com/component/mobile-id/}}
\footnotetext[2]{Finnish Mobile Certificate:
\url{http://www.mobiilivarmenne.fi/en/}}

Given the centrality of the role eID architectures will play in the
lives of people, it is important to identify and ensure the security,
usability and regulatory requirements an eID architecture should meet
in order to be deemed trustworthy. There are two major requirements
that need to be met. First, the link between the natural person
represented by an eID and the person using that eID must be
established. Legislation typically provides exact regulation on the
required issuance procedure for physical identity tokens. Since an eID
is to be interchangeable with a physical one, the same laws need to be
applied to eID issuance with little or no modification~\cite{Myhr08}.
Second, the eID credentials must be logically isolated from other
software and must be used only when the person associated with the eID
intends it. The latter requirement is met by subjecting the use of the
eID credential to user authentication, usually based on PINs. PINs
protecting eID credentials are subject to a number of policy
restrictions intended to balance security and
usability~\cite{Laitinen14}. For instance, the eID issuer or a service
provider relying on eID may stipulate the type of protection (e.g.
PIN) for the credential. Also, since a personal device can hold
multiple eID-related credentials, it must be possible to associate the
same PIN with multiple credentials.

The Trusted Computing Group (TCG)\footnotemark\ is the leading
organization specifying standards for trusted hardware on a wide range
of devices ranging from servers and personal computers to mobile
devices. Hundreds of millions of \emph{Trusted Platform Modules} (TPMs)
are deployed in such devices.  The new version of the TPM
specification\footnotemark\ includes a new, rich authorization model
for specifying flexible access control policies for objects protected
by the TPM\@. This makes it possible to support the kind of complex
policies required for protecting eID credentials on the range of
devices traditionally equipped with TPMs.  TPM 2.0 specifications have
been public for some time. However, as far as we are aware, there has
been no accessible explanation of the TPM 2.0 authorization model.

\footnotetext[3]{\url{http://www.trustedcomputinggroup.org/}}
\footnotetext[4]{\url{http://www.trustedcomputinggroup.org/media_room/news/352}}

In this paper, we describe how an eID architecture can be designed and
implemented using TPM 2.0. In particular, we make the following
contributions:
\begin{itemize}
\item We provide the \textbf{first accessible description of the TPM
  2.0 authorization model}. We believe that this will help researchers
  understand the power of the model and to design other security and
  privacy solutions based on it.
\item We describe the \textbf{detailed design of an eID architecture} 
  based on TPM 2.0. We show how this solution improves the overall
  security and usability compared to eID solutions based on traditional
  physical tokens such as removable smart cards.
\item We identify \textbf{possible enhancements to the TPM 2.0
  authorization model}. If adopted, these enhancements would further
  improve the flexilibility of the authorization model, and greatly
  simplify the definition of certain kinds of policies.
\end{itemize}

The paper is organized as follows: Section~\ref{sec:background}
introduces necessary background on current European eID solutions and
TPMs. In particular, Section~\ref{sec:tpm2-extended-authorization}
describes the TPM 2.0 authorization model in detail.
Section~\ref{sec:requirements} introduces the requirements for our
design, while the design itself is presented in
Section~\ref{sec:design}. An analysis of the design is given in
Section~\ref{sec:analysis}. Section~\ref{sec:perceived-ea-shortcomings}
suggests some potential improvements to the current TPM 2.0
authorization model. Related work is discussed in
Section~\ref{sec:related-work}. Section~\ref{sec:conclusion} concludes.

\setlength{\textfloatsep}{.05cm}

\section{Background}
\label{sec:background}

\subsection{eID in Europe}

The legal status of eID mechanisms in Europe is based on the EU signature
directive from 1999 (1999/93/EC) and the data protection and privacy directives
from 1995 (95/46/EC) and 2002 (2002/58/EC)~\cite{Mahler13}. These directives
guarantee legal equivalence between digital signatures done with the citizen
cards and physical signatures by the associated person. They set the privacy
framework for how identity information can be stored and managed in servers
interacting with identified citizens.

Technically, the current European ID cards are often (contactless) ISO 7816
smart cards, corresponding to the PKCS\#15 interface standard of which there are
jurisdiction-specific profiles. Most identity cards contain at least two
signature keys and provide two forms of signatures --- the legally binding
“qualified” signature key for document signing is separated from the
“authentication” keys to be used in more interactive settings. Key usage
authorization is governed with PINs.

Such identity tokens come with a number of drawbacks that influence large-scale
deployment; the physical token must be present for each authentication, and all
devices must be equipped with compatible card readers. As most smart cards lack
user interfaces, if credential authorization requires a PIN, users typically
have to enter it via a potentially untrusted device.

Nevertheless, in the EU, the main challenges hindering the adoption of an
interoperable pan-European eID scheme are considered to be not only
technological, but also legal. In a recent report by the eID team in the
European Commission's Institute for Prospective Technological Studies
(IPTS)~\cite{Andrade13}, a series of principles that aim at providing the legal
underpinning for a future eID legal scheme are elaborated on. These include,
among others, the principles of user-centricity, anonymity and pseudonimity and
the principles of multiple identities, identity portability and un-linkability.

\subsection{FIDO Alliance}

\begin{figure}[tb]
  \centering
  \includegraphics[width=0.8\linewidth]{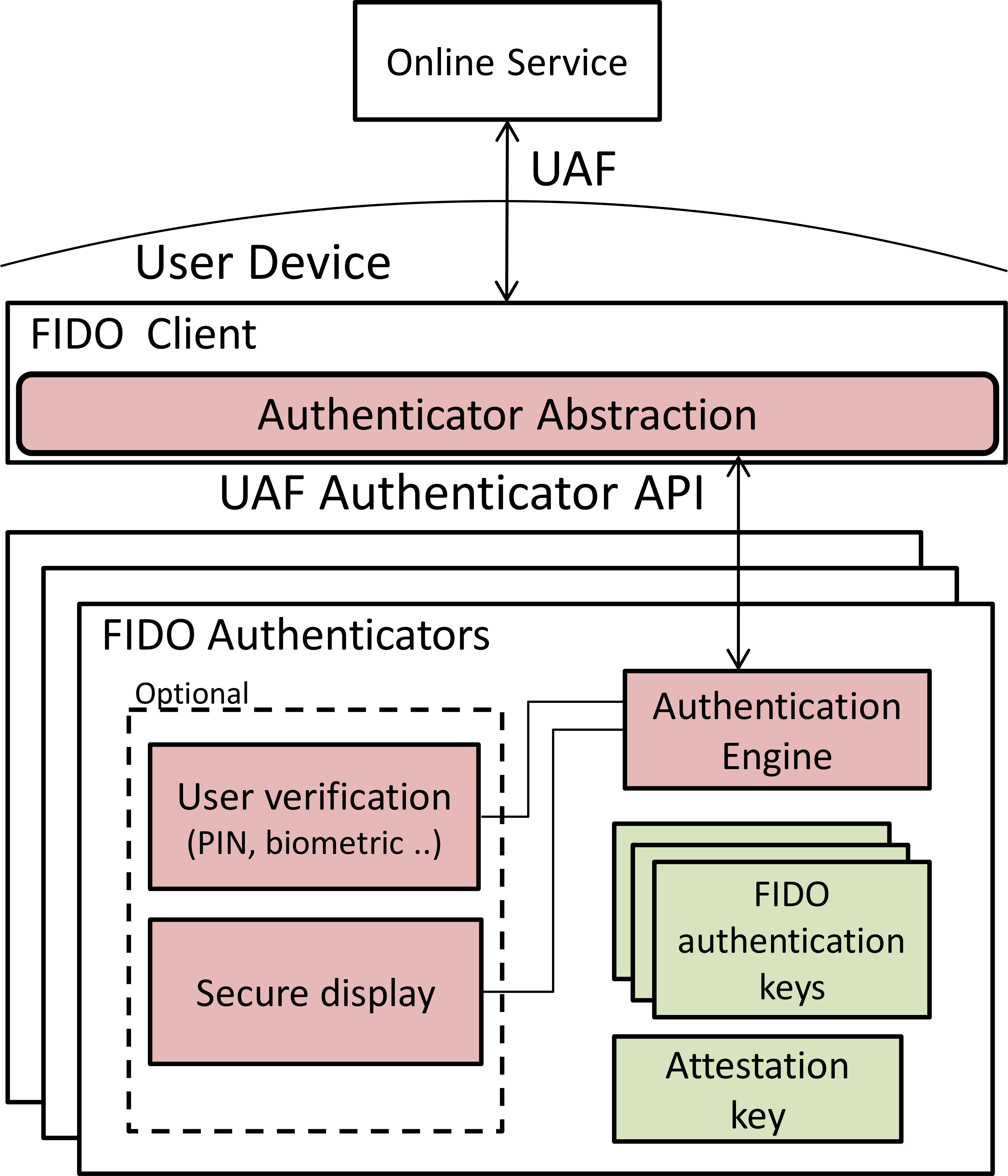}
  \caption{FIDO authenticator reference model}
  \label{fig:fido}
\end{figure}

The Fast Identity Online (FIDO) alliance\footnotemark\ is an industry consortium
whose mission is to reduce the reliance on passwords for user authentication in
internet transactions. The alliance has the backing of many industry giants, and
may turn out to be a significant reference model even for government eID
applications.

\footnotetext{\url{https://fidoalliance.org/specifications}}

The FIDO specifications define authentication architectures and protocols based
on strong authentication and biometrics~\cite{Reimer13}. The \emph{Universal
Authentication Framework} (UAF) protocol enables online services to support
password-less and multi-factor security. In fact, a typical eID credential is
very close to the reference authenticator component in the FIDO architecture
(see Figure~\ref{fig:fido}).  The credential can (and should) have support for
device-local user authentication for credential enablement, that can take the
form of a PIN, password or e.g.\ a biometric.

However, what sets the FIDO protocols apart from many legacy eID setups, is that
FIDO clients generate service-provider specific asymmetric key pairs for
authentication. Using a different key pair with every service provider improves
user privacy, since it protects against linking the activity of a user in cases
where service providers collude. Supporting many dynamically allocatable
authentication keys is not technically difficult to implement in TPMs, or even
with smart cards. This is a practical requirement also for government eID. In
such a set up, it is desirable to bind several key pairs to the same instance of
device-local user authentication.

\ifdefined\istechreport
For mobile devices, and for the use cases discussed in this paper, the FIDO
token primarily operates as a \textit{First Factor Bound
Authenticator}~\cite{Lindeman14}. In this mode of operation, the primary means of
authentication is strengthened by the authenticator being bound to an
identifiable / attestable device. In the FIDO architecture, a credential can
also be removable, either as a \textit{First Factor Roaming Authenticator}, or
augmenting another authentication as a \textit{Second Factor Authenticator}.
\fi

\subsection{TPMs}

TPMs are system components that provide a hardware-based approach for secure
\emph{non-volatile} (NV) storage, cryptographic key generation and use, sealed
storage and (remote) attestation. The primary scope of a TPM is to assure the
integrity of a platform by providing means to identify (and report on) the
hardware and software components that comprise the platform. The notion of
\emph{``trust''} in the context of TPMs stems from the expectation of behaviour
that can be determined from this identity. The state of the TPM is separate from
the state of the platform on which it reports, and the only way for the host
system to interact with a TPM is through an interface defined in the TPM
specifications~\cite{TPM1.2,TPM2.0}. TPMs differentiate themselves from
conventional secure cryptoprocessors in that they provide \emph{platform
binding}, i.e.\ proof of an association between a cryptographically verifiable
identity and the platform itself.

TPMs can be implemented as single-chip components with separate physical
resources dedicated to the TPM\@. The first TPM microcontrollers became
available in 2005\footnotemark, and since then TPMs are included in many new
laptop computers, primarily in business product lines. However, the notion of a
platform in the context of TPMs is not tied PCs, nor to a particular operating
system. Hardware-based \emph{Trusted Execution Environments}~\cite{Ekberg14}
(TEEs) have become commonplace in modern mobile platforms. They provide secure,
integrity-protected processing environments, consisting of processing, memory
and storage capabilities, that are isolated from the regular processing
environment, sometimes called the \emph{rich execution environment} (REE), where
the device operating system and applications run. As TEEs provide the necessary
capabilities for housing TPMs, firmware-based TPMs are a more likely deployment
model on mobile devices.

\footnotetext{Infineon press release announcing the SLB 9635 TPM 1.2:\newline
\url{http://www.infineon.com/cms/en/corporate/press/news/releases/2005/132443.html}}

\subsubsection{Attestation}
\label{sec:attestation}

Compared to smart cards, the issuance of secure functionality to a TEE in a
mobile device out in the field will require an extra set of trust anchors. This
particular problem setting is described in a whitepaper on \emph{User-Centric
Provisioning}~\cite{GlobalPlatform12} published by the GlobalPlatform
association\footnotemark. Compared to physical smart cards, which as a rule are
provisioned by a trustworthy manufacturer on behalf of the token issuer, a TPM
as a TEE function or application is manufactured and put to market in a trust
domain that normally does not \emph{a priori} include any relation to the
service issuer. Thus, the connection between the token issuer trust domain and
the TEE must be established at service provisioning time.

\footnotetext{\url{http://www.globalplatform.org/}}

At large this problem is one of \textit{attestation}, i.e.\ the service issuing
entity must both authenticate the targeted endpoint as well as convince itself
of the current setup, version and revocation status of the TEE deployed in that
endpoint. Furthermore, for data provisioning the issuer must confirm that the
endpoint within the TEE, i.e.\ the \emph{Trusted Application} (TA), will be the
issued service and not some other code running in the TEE\@.

In physical TPMs, the attestation property is embodied by \emph{Endorsement
Keys} (EKs) — unique device secret keys that are associated with a device
certificate via a trusted party, like the TPM manufacturer. In the TPM 1.2
specification the endorsement key is a decryption key which, using a trustworthy
protocol to a “privacy CA”, can certify locally generated \emph{“Attestation
Identity Keys”} (AIKs) – signature keys used in the actual attestation events.
In the TPM 2.0 specification the endorsement key can also be a signature key.
With physical TPM chips, the generation of the EK credential can take place
during chip manufacturing, and the credential is best stored in the non-volatile
memory of the TPM chip for later use. This approach needs no attestation
external to the TPM, since the manufacturing process can be secured to a needed
level. On the other hand, for post-manufacturing deployment of firmware TPMs,
relevant attestation needs to be provided by the underlying platform. In this
paper we assume that the TPM 2.0 (and its EK) can be properly attested by an
underlying functionality, and therefore the eID service provider, whether being
the government or a private entity, can trust the TPM they are interacting with
as the device-specific, fully isolated security component that it is defined to
be.

\subsubsection{PCRs}

Platform Configuration Registers (PCRs) are integrity-protected registers in
TPMs that are used to store aggregate measurements regarding the security state
of the system. A PCR value is a representation of the state of a particular
(software) environment. Each PCR holds a digest value consisting of an
accumulative hash of previous PCR values.  Apart from resetting the PCR to an
initial value, the only way to modify a PCR value is to \emph{extend} a
measurement value into the PCR\@. When a measurement is extended into a PCR, the
new digest value is calculated as follows:

\vspace{.5\baselineskip}
\noindent
$PCR_{new} = \mathbf{H}(PCR_{old} || digest)$

\vspace{.5\baselineskip}
\noindent where

\vspace{.5\baselineskip}
\begin{description}
\vspace{-.25cm}
  \scriptsize
  \item[$PCR_{new}$] \hfill is the new digest value being stored in the PCR
  \item[$PCR_{old}$] \hfill is the previous digest value stored in the PCR
  \item[$\mathbf{H}()$] \hfill is the hash function associated with the PCR
  \item[$digest$] \hfill is the measurement value extended into the PCR
\end{description}

As a result of the old digest value being hashed into the new one, any deviation
in a reported sequence of events causes an irrevocable change in the eventual
PCR digest value. In other words, the PCR digest value is unique for the
specific order and combination of digest values that have been extended into a
particular PCR\@. The state of a system, represented by a set of PCRs, can be
used in attestation (see Section~\ref{sec:attestation}).

\vspace{-.5\baselineskip}

\subsubsection{TPM 1.2 Authorization}

\begin{figure*}
  \centering
  \begin{subfigure}{.5\textwidth}
    \centering
    \includegraphics[width=\textwidth]{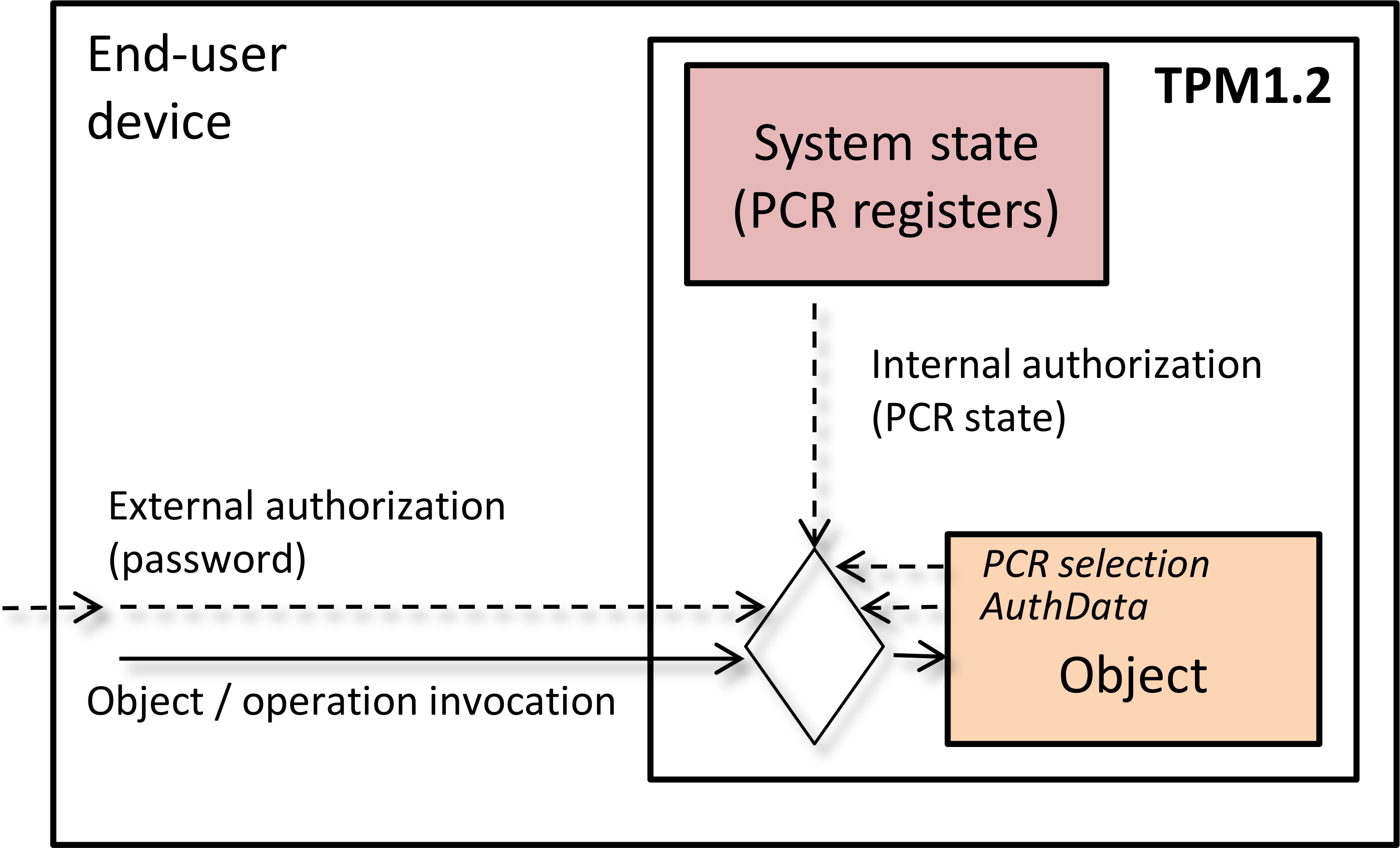}
    \caption{TPM 1.2 authorization model}
    \label{fig:tpm-authorization}
  \end{subfigure}%
  \begin{subfigure}{.475\textwidth}
    \centering
    \includegraphics[width=\textwidth]{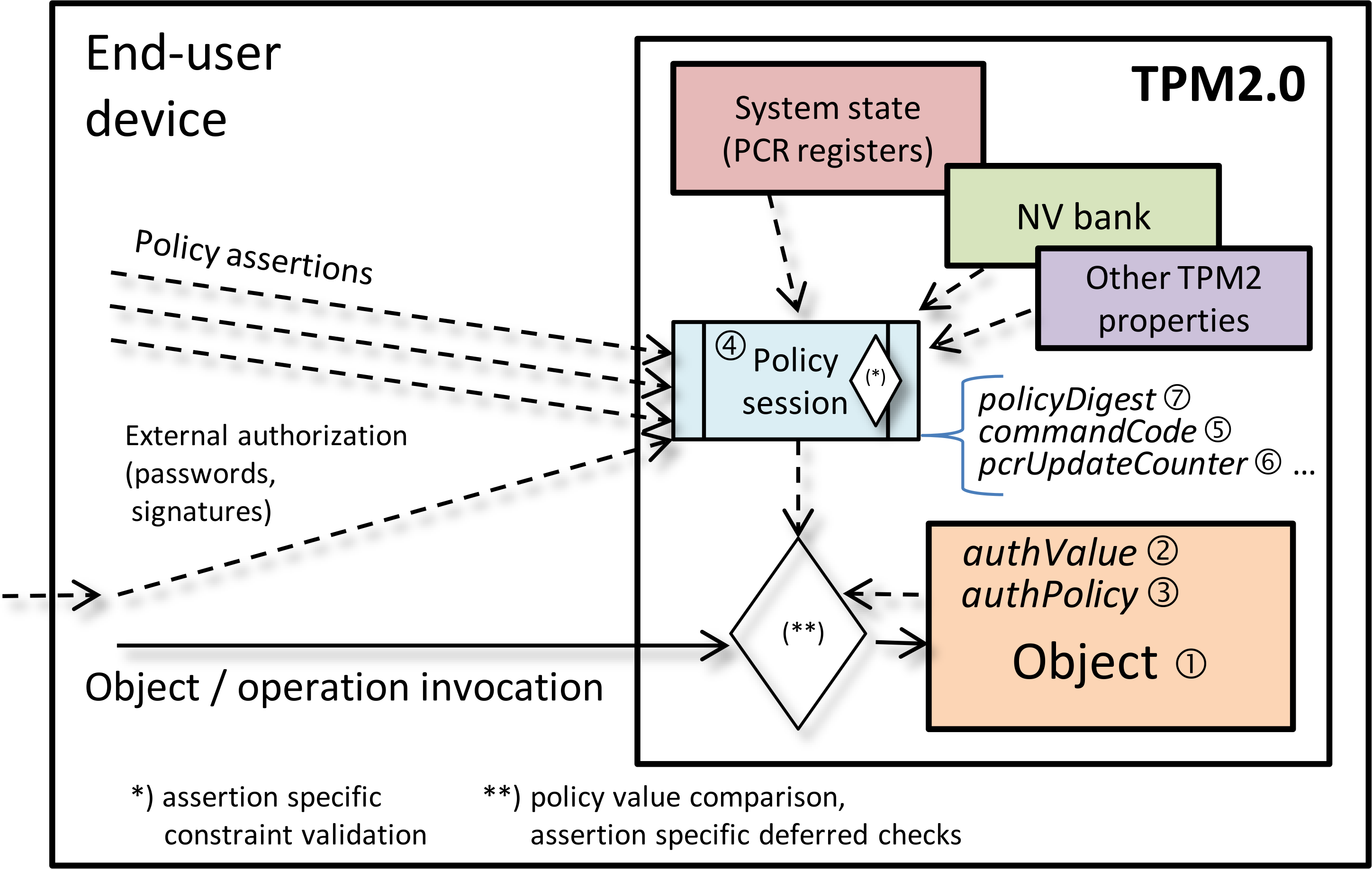}
    \caption{TPM 2.0 enhanced authorization model}
    \label{fig:tpm2-authorization}
  \end{subfigure}
  \caption{Overview of TPM authorization models}
  \label{fig:tpm-authorization-models}
\end{figure*}

In the TPM 1.2 specification~\cite{TPM1.2}, access to TPM operations and objects
secured by the TPM (e.g.\ cryptographic keys) are protected via an authorization
mechanism. Access to such an object is obtained via the proof of knowledge of a
shared secret associated with the object. In TPM parlance, this shared secret is
known as the \texttt{AuthData} of the object. Henceforth, we refer to this
shared secret as the \emph{owner authorization value}, as in general, knowledge
of the \texttt{AuthData} is treated as complete proof of ownership of a
protected object or operation, with the exception of asymmetric keys locked to a
set of particular PCR values during their creation (as discussed below). The TPM
places no additional requirements on the use of the object. An overview of the
TPM 1.2 authorization model is shown in Figure~\ref{fig:tpm-authorization}.

In order to securely pass proof-of-knowledge of the owner authorization value to
the TPM, three protocols are used; the \emph{Object-Independent Authorization
Protocol} (OIAP), the \emph{Object-Specific Authorization Protocol} (OSAP), and
the \emph{Delegate-Specific Authorization Protocol} (DSAP). These protocols
allow the caller to establish a confidential authorization session with the
TPM\@. Depending on the protocol used, the authorization session has different
properties, e.g.\ the OIAP protocol allows access to multiple protected objects
to be authorized during the same authorization session, with the limitation that
OIAP sessions cannot be used for key creation, or other operations that would
introduce new authorization information to the TPM\@. OSAP sessions, on the
other hand can only be used manipulate a single object, but allows new
authorization information to be transmitted to the TPM\@. DSAP provides support
for delegating access to an object without disclosing its owner authorization
value. Instead, object owners may specify a set of operations on an object,
which are authorized via an \emph{delegation authorization value}, provided by
the caller as the \texttt{authData} instead of the owner authorization value
when operations on an object are invoked using delegated privileges.
\ifdefined\istechreport For further information on the TPM 1.2 authorization
protocols, we refer the reader to Appendix~\ref{appendix:tpm-authprotocols}. \fi

In addition, as mentioned earlier, cryptographic keys may be locked to
particular PCR values upon key creation. If this is the case, the key is only
usable as long as certain PCRs have the particular values associated with the
locked key. In this way, PCR values can be used to ensure that certain keys are
accessible only to authorized software. This is typically used in combination
with the DSAP delegation mechanism to allow a trusted process access to a
protected object without user intervention. DSAP checks for the continued
validity of such PCR selections, and any change to the PCR values causes the
invalidation of the DSAP session. The combination of the delegate authorization
value and PCR selections gives the TPM 1.2 four distinct authorization modes,
listed in Table~\ref{tbl:tpm-authmodes}. When a PCR selection is set, the
delegation authorization value associated with the delegated key may be a fixed,
well-known, value (case~\textcircled{3}). This is because, if the trusted
process is to execute automatically, the only way for the trusted process to
store the delegation authorization value would be to \emph{seal}\footnotemark\
it against the process's PCR measurement values, but as the delegation mechanism
already checks the PCR selections, the verification of the delegate
authorization value is redundant.

\footnotetext{Sealing data against some PCR values refers to the act of
  encrypting the data in such a way that the TPM will later decrypt it only if
  the PCRs have the given values. This may be used to ensure that the data can
  be opened only if the platform is in a known and trusted state.}

\begin{table}
  \begin{tabular}{| c | c | c | c |}
    \hline
    case & \texttt{authValue} & PCR selection & authorization method\\ \hline
    \scriptsize\textcircled{1} & secret     & yes & \pbox{20cm}{password \emph{and} \\ platform state} \\
    \scriptsize\textcircled{2} & secret     & no  & password  \\
    \scriptsize\textcircled{3} & well-known & yes & platform state \\
    \scriptsize\textcircled{4} & well-known & no  & ---       \\ \hline
  \end{tabular}
  \caption{TPM 1.2 authorization modes}
\label{tbl:tpm-authmodes}
\end{table}

\subsubsection{TPM 2.0 Enhanced Authorization}
\label{sec:tpm2-extended-authorization}

\begin{table*}
  \begin{center}
  \small
  \begin{tabular}{| l  l | p{0.5\textwidth} |}
    \hline
    \multicolumn{1}{| l }{} & \multicolumn{1}{ c |}{Item} & \multicolumn{1}{ c |}{Description} \\ \hline
    \scriptsize\textcircled{1} & $Object$ & Key or data stored within the TPM \\
    \scriptsize\textcircled{2} & $Object \rightarrow authValue$ & Byte string used as password in access authorization \\
    \scriptsize\textcircled{3} & $Object \rightarrow authPolicy$ & Digest value used for policy session access authorization \\
    \scriptsize\textcircled{4} & $policySession$ & Authorization used for gaining access to an $Object$ by satisfying an associated policy \\
    \scriptsize\textcircled{5} & $policySession \rightarrow commandCode$ & Command code for TPM command being authorized \\
    \scriptsize\textcircled{6} & $policySession \rightarrow pcrUpdateCounter$ & Stored PCR update counter reading \\
    \scriptsize\textcircled{7} & $policySession \rightarrow policyDigest$ & Digest value calculated as a result of policy command invocation during a policy session \\
    \hline
  \end{tabular}
  \end{center}
  \caption{Glossary of TPM 2.0 Enhanced Authorization Terminology}
\label{tbl:tpm2-glossary}
\end{table*}

In the TPM 2.0 specification~\cite{TPM2.0}, objects stored within the TPM have
an \texttt{authValue} property associated with them. The \texttt{authValue} is
directly comparable to the TPM 1.2 \texttt{AuthData}, and may be used as a
password for object authorization. In addition, the TPM 2.0 specification
introduces \emph{Enhanced Authorization} (EA) policies, which supersede the TPM
1.2 authorization mechanisms and allows object owners and administrators to
require specific assertions or actions to take place before access to a
protected object is allowed. The policy associated with an object may be
arbitrarily complex, even though internally the policy is reduced to a single
statistically unique digest value known as the \texttt{authPolicy}. The
\texttt{authPolicy} is associated with the TPM entities the corresponding policy
applies to. An overview of the TPM 2.0 enhanced authorization model is shown in
Figure~\ref{fig:tpm2-authorization}.

Access to all objects making use of enhanced authorization takes place via a
session-based authorization procedure. In order to access a TPM object, the
caller initiates a policy session with the TPM\@. Subsequently the caller issues
a sequence of \emph{policy commands} to the TPM.

Each policy command in the authorization policy is an assertion that a
particular statement is true in order for the policy to be satisfied. For
instance, a particular policy might require that certain PCRs have specific
values in order for access to a TPM object to be authorized. As a side-effect of
a true policy assertion, each such policy command modifies a digest value
associated with the session, characteristic of the particular policy expressed
via the sequence of policy commands. This running accumulation of the digest
value is called the \texttt{policyDigest}.  When a policy session is started,
the associated \texttt{policyDigest} is initialized to zero. Then, as each
assertion corresponding to a certain policy command is evaluated, the
\texttt{policyDigest} is updated in a manner similar to PCR extension:

\vspace{.5\baselineskip}
\noindent
\resizebox{\linewidth}{!}{%
  $policyDigest_{new} = \mathbf{H}(policyDigest_{old}\;||\;commandCode\;||\;commandArgs)$
}

\vspace{.5\baselineskip}
\noindent where

\begin{description}
  \small
  \item[$policyDigest_{new}$] \hfill is the new policy session digest value
  \item[$policyDigest_{old}$] \hfill is the previous policy session digest value
  \item[$\mathbf{H}()$]       \hfill is the hash function used to update the \texttt{policyDigest}
  \item[$commandCode$]        \hfill is a value which identifies the policy command
  \item[$commandArgs$]        \hfill are dependent on the condition asserted
\end{description}

Some policy commands also have the ability to reset the \texttt{policyDigest}
value. This occurs conditionally with regard to the previous
\texttt{policyDigest} value. We will see that this construct enables not only
branching policies, but also policies to be changed based on signatures
generated by an external authorization entity:

\vspace{.5\baselineskip}
\noindent
\resizebox{\linewidth}{!}{%
  $\begin{aligned}
    \mathbf{if}&\;condition\;\mathbf{then} \\
    &\;\;policyDigest_{new} = \mathbf{H}(0\;||\;commandCode\;||\;commandArgs)
  \end{aligned}$
}
\vspace{.5\baselineskip}

Only policy commands modify the \texttt{policyDigest}; other TPM 2.0 commands do
not. Later in this section, we provide selected examples of different types of
policy commands. \ifdefined\istechreport For descriptions of other commands
referred to elsewhere in this paper, we refer the reader to
Appendix~\ref{appendix:extended-authorization-commands}. \else For descriptions
of other commands referred to elsewhere in this paper, we refer the reader to
our accompanying technical report~\cite{Nyman14}.
\fi

Finally, after the policy command sequence has been completed, the final value
of the \texttt{policyDigest} for the session is compared to the
\texttt{authPolicy} of the object being accessed. A match indicates that the
sequence of invoked policy commands matches, and satisfies the assertions
expressed by the policy, successfully completing the authorization. The
\texttt{authPolicy} value associated with an object upon creation can either be
calculated in software, or in a special \emph{trial policy session}, during
which all assertions are assumed to succeed and the \texttt{policyDigest} can be
retrieved from the TPM at the end of the session.

\paragraph{Types of Policy Assertions}

TPM 2.0 EA policy commands fall into three categories: \emph{immediate},
\emph{deferred}, and \emph{combined} assertions. In this section, we will
provide examples and discuss the specifics of each of the three policy
assertion types.

\noindent\textbf{Immediate assertions} are policy commands which only affect the
\texttt{policyDigest}. An example of an immediate policy assertion is the
\texttt{TPM2\_PolicyNV()} command, which asserts an arithmetic comparison
between an input value and a value stored in a specified non-volatile storage
element. If the condition holds, the \texttt{policyDigest} is updated
accordingly, otherwise it remains unchanged. Specifically:

\vspace{\baselineskip}
\noindent
\resizebox{\linewidth}{!}{%
    $\begin{aligned}
      \mathbf{if}&\;NValue\;\mathbf{op}\;operand\;\mathbf{then} \\
      &\;\;policyDigest_{new} = \mathbf{H}(policyDigest_{old}\;||\;TPM\_CC\_PolicyNV\;||
      \;args\;||\;NVName)
    \end{aligned}$
}

\vspace{.5\baselineskip}
\noindent where

\begin{description}
  \small
  \item[$TPM\_CC\_PolicyNV$]  \hfill is the policy command code
  \item[$args$]    \hfill is a hash over the input operands and operator
  \item[$NVValue$] \hfill is the value of the non-volatile storage element
  \item[$NVName$]  \hfill is the name of the non-volatile storage element
\end{description}

\noindent\textbf{Deferred assertions} unconditionally update the
\texttt{policyDigest} based on input values and record specific constraints in
the context of the current policy session. When the session is used to make the
final authorization decision, the stored constraints are validated at that time.
For instance, the \texttt{TPM2\_PolicyCommandCode()} policy command is used to
verify that the policy session is only used to authorize a particular command.
This is achieved by storing the command code for the command being authorized in
the current session context. The \texttt{policyDigest} is updated
unconditionally, namely:

\vspace{.5\baselineskip}
\noindent
\resizebox{\linewidth}{!}{%
  $policyDigest_{new} =
  \mathbf{H}(policyDigest_{old}\;||\;TPM\_CC\_PolicyCommandCode\;||
  \;code)$
}

\vspace{.5\baselineskip}

where $code$ is the command code for the TPM command being authorized. When the
final authorization decision is made, the TPM will verify that the command used
to operate on the object being authorized is in fact the command identified by
the command code stored in the session context.

\noindent\textbf{Combined assertions} validate a precondition regarding the
TPM state, \emph{and} record some parameters in the current policy session
context used for deferred checks later on. An example is the
\texttt{TPM2\_PolicyPCR()} command, which can be used to validate that a
specified PCR has the expected value. If the caller provides an expected value,
the value of the specified PCR is compared immediately to the expected value; if
the values match, the \texttt{policyDigest} is updated accordingly:

\vspace{.5\baselineskip}
\noindent
\resizebox{\linewidth}{!}{%
  $policyDigest_{new} =
  \mathbf{H}(policyDigest_{old}\;||\;TPM\_CC\_PolicyPCR\;||
  \;PCRs\;||\;digest)$
}

\vspace{.5\baselineskip}
\noindent where

\begin{description}
  \small
  \item[$PCRs$]   \hfill is a bit mask corresponding to the PCR selection
  \item[$digest$] \hfill is the hash of the PCR values in the selection
\end{description}

If the caller does not provide an expected value, the \texttt{policyDigest} is
updated as indicated above, but the validity of the PCR values will not be known
until the policy session is used for authorization (i.e.\ when the
\texttt{policyDigest} is compared to the \texttt{authPolicy}).  However, merely
verifying the PCR value as part of a precondition leaves the authorization
policy susceptible to \emph{Time-Of-Check Time-Of-Use}\footnotemark\ (TOCTOU)
race conditions in cases where PCR values have changed in the interval between
the invocation of the \texttt{TPM2\_PolicyPCR()} command, and the time the actual
authorization decision is made. To avoid TOCTOU conditions, the TPM keeps track
of PCR changes by incrementing a monotonically increasing counter, the
\texttt{pcrUpdateCounter}, each time a PCR is updated.\ When
\texttt{TPM2\_PolicyPCR()} is invoked, the current value of the PCR update
counter is stored in the current session context. If an expected value was
provided, the PCR update counter is updated only if the immediate PCR value
check succeeds. On subsequent \texttt{TPM2\_PolicyPCR()} invocations, and when
the policy session is used for authorization, the value of the PCR update
counter is compared against the stored counter; the authorization will fail
unless the counter values match.

\footnotetext{CWE-367: Time-of-check Time-of-use Race Condition:\newline
\url{http://cwe.mitre.org/data/definitions/367.html}}

\paragraph{Policy OR, and AND}
\label{sec:policy-or}

As mentioned in Section~\ref{sec:tpm2-extended-authorization} branching policies
are made possible by resetting the \texttt{policyDigest} value as part of the
successful assertion. This disrupts the running digest value, allowing
subsequent assertions to proceed from an independent, yet well-known, value. The
logical disjunction is embodied by the \texttt{TPM2\_PolicyOR()} command. When
invoked, \texttt{TPM2\_PolicyOR()} is passed a list of digest values, each
corresponding to a digest value that is accepted as a valid precondition for a
successful \texttt{TPM2\_PolicyOR()}. The \texttt{policyDigest} is updated
conditionally only if its current value is in this list:

\vspace{.5\baselineskip}
\noindent
\resizebox{\linewidth}{!}{%
  $\begin{aligned}
    \mathbf{if}&\;policySession \rightarrow
    policyDigest\;\mathbf{in}\;digest_{1} \ldots digest_{n}\;\mathbf{then} \\
    &\;\;policyDigest_{new} =
    \mathbf{H}(0\;||\;TPM\_CC\_PolicyOR\;||\;digests)
  \end{aligned}$
}

\vspace{.5\baselineskip}
\noindent where

\begin{description}
  \small
  \item[$digest_{1} \ldots digest_{n}$] \hfill is a list of valid digest values\footnotemark
  \item[$digests$] \hfill is the concatenation $digest_{1}\;||\;\ldots\;||\;digest_{n}$
\end{description}

\footnotetext{We note that the TPM 2.0 specification limits the number of
digests to $n=8$. However, by nesting multiple \texttt{TPM2\_PolicyOR()}
operations, the effective size of the list can be expanded indefinitely.}

The reasoning behind this scheme is that
$\mathbf{H}(digest_{1}||\ldots||digest_{n})$ is a well-known, fixed value. For
an arbitrary $digest' \notin \{digest_{1} \ldots digest_{n}\}$, it is
computationally infeasible, due to the properties of the hash function
$\mathbf{H}$, to find a concatenation
$digest_{x}||\ldots||digest'||\ldots||digest_{z}$ such that:

\vspace{.5\baselineskip}
\noindent
\resizebox{\linewidth}{!}{%
  $\mathbf{H}(digest_{1}||\ldots||digest_{n}) =
  \mathbf{H}(digest_{x}||\ldots||digest'||\ldots||digest_{z})$
}

\vspace{.5\baselineskip}

Although the set of EA policy commands does not include an explicit logical
\textbf{AND} operation, the way the \texttt{policyDigest} is updated by policy
commands, each new value being dependent on the previous one, not only acts as
an implicit logical conjunction, but also imposes an order dependence on the
sequence of consecutive policy commands.

\paragraph{External Authorization}
\label{sec:external-authorization}

While some EA policies within the TPM may be altered by changing the
corresponding \texttt{authPolicy} value, policies associated with key objects
and NV memory elements may not. However, it is often the case that static access
control policies are \emph{``brittle''} in the sense that they cannot
accommodate for changes in the system that makes the policy essentially
unusable. This applies for instance to a policy that seals a piece of data to a
set of fixed PCR values to ensure that the data is only accessible when the
system has been booted in a specific configuration. In this case, the PCR values
could correspond to measurements of the BIOS and operating system kernel. If
there is a BIOS update by the Original Equipment Manufacturer (OEM), the
measurement value in the corresponding PCR will change, rendering the data
inaccessible because the policy cannot accommodate for the BIOS update in
advance. These kinds of situations require flexible policies that may be
modified in an indirect way after-the-fact. For such cases, the TPM 2.0 EA
mechanism provides the \texttt{TPM2\_PolicyAuthorize()} command.

The \texttt{TPM2\_PolicyAuthorize()} command asserts that the current
\texttt{policyDigest} is authorized by an external entity via the signing of the
corresponding digest value and an optional policy classifier\footnotemark. If
the \texttt{policyDigest} is authorized in such a way, it is reset, and replaced
by the \emph{name} of the signing key used for authorization. If present, the
policy classifier can act as a nonce to limit the use of the signature key, as the
classifier is extended to the \texttt{policyDigest}:

\vspace{.5\baselineskip}
\noindent
\resizebox{\linewidth}{!}{%
  $\begin{aligned}
    \mathbf{if}&(\mathbf{V}(keyHandle_{A}, signature, policySession \rightarrow policyDigest)\;\\
    &\;\;policyDigest_{new} = \mathbf{H}(0\;||\;TPM\_CC\_PolicyAuthorize\;||\;keyName_{A})\\
    &\;\;policyDigest_{new+1} =\mathbf{H}(policyDigest_{new}\;||\;policyRef)
  \end{aligned}$
}

\vspace{.5\baselineskip}
\noindent where

\begin{description}[itemsep=.1cm]
  \small
  \item[$A$]                 \hfill is the authorization key
  \item[$\mathbf{V}()$]      \hfill is the signature verification algorithm for $A$
  \item[$keyHandle_{A}$]     \hfill is the handle for $A$
  \item[$signature$]         \hfill is the signature of the policy to be authorized
  \item[$keyName_{A}$]       \hfill is the name of the object corresponding to $A$
  \item[$policyRef$]         \hfill is the policy classifier
\end{description}

\footnotetext{In practice, the signature is verified by the TPM in a separate
  \texttt{TPM2\_VerifySignature()} command invocation, which produces a ticket
  that is subsequently passed to \texttt{TPM2\_PolicyAuthorize} to, if valid,
  provide proof that the TPM has validated the signature using a particular key.
  The reason for this seems to be purely a performance optimization, as it is
  more efficient to verify the ticket than it would be to load a TPM object each
  time authorization occurs to verify the signature.}

The public portion of key objects stored by the TPM is represented in the name
of the object. In the case of asymmetric keys, this means that the new
\texttt{policyDigest} value is derived, in part, from the public key of the
authorizing entity. If a symmetric key is used, its public portion is derived by
hashing the key material together with obfuscation values which prevents the
public portion from leaking information about the sensitive key material. In
other words, any key can be used for the external authorization, but each
individual key will leave a matching trace in the \texttt{policyDigest}.
Therefore that binding will constitute the identification of the key needed for
the authorization.

Returning to our sealing example, the policy that seals the data can be made
robust with regards to software updates via a \texttt{TPM2\_PolicyAuthorize()}
that occurs after the PCR measurements have been extended into the
\texttt{policyDigest} with \texttt{TPM2\_PolicyPCR()}\footnotemark. This allows
the OEM to provide a signature for PCR values corresponding to the new BIOS\@.
With either authorized set of of PCR values, the final \texttt{policyDigest}
value will be the same, even though the intermediate values after
\texttt{TPM2\_PolicyPCR()} differ.

\footnotetext{In this case, the expected PCR values would not be provided by the
caller in the \texttt{TPM2\_PolicyPCR()} invocation.}

\section{Requirements}
\label{sec:requirements}

The requirements for an eID token can be divided into \textit{functional},
\textit{security}, and \textit{privacy} requirements. Some translate into
\textit{platform requirements}, others are restricted to the token itself or
apply to token interaction. An overview of
eID requirements is shown below:

\begin{center}
\begin{enumerate}
  \item \label{req:functional-requirements} Functional requirements:
  \begin{enumerate}
    \item \label{req:identity-binding} Binding the identity of a physical person to
    multiple identity keys stored on personal device
    \begin{enumerate}
      \item \label{req:one-to-many} One-to-many relationship between PIN and keys
      \item \label{req:pin-and-puk} PIN and PUK with limited number of tries
    \end{enumerate}
    \item \label{req:api} API compatibility with smart card-based system
  \end{enumerate}
  \item \label{req:security-requirements} Security requirements:
    \begin{enumerate}
      \item \label{req:cryptograpy} Cryptographic requirements
      \item \label{req:confidentiality} Confidentiality of identity key
      \item \label{req:isolation} Code isolation of operations on the key
      against software and other keys stored on the device
    \end{enumerate}
  \item \label{req:privacy-requirements} Privacy requirements:
  \begin{enumerate}
    \item \label{req:authorization} Prevention of unauthorized access to credential
    \item \label{req:linkability} Linkability control
  \end{enumerate}
\end{enumerate}
\end{center}

\subsection{Functional requirements}

The \emph{Application Programming Interface} (API) to eID tokens resident in
mobile devices must be as similar as possible to the API of existing smart
card-based ID systems (Req.~\ref{req:api}), i.e.\ a PKCS\#15-based interface,
with a 3-attempt PIN and a PUK for enabling key signatures
(Req.~\ref{req:pin-and-puk}) is the baseline in such an ecosystem. For TPM-based
eID, a compatibility layer to provide the necessary API interoperability is
needed.

Application requirements for eID sometimes also support and refine the general
security requirements. In discussions with representatives of our local
jurisdiction (personal correspondence, Feb 28, 2014)~\cite{Laitinen14}, it became
clear that there is a wish to address the binding problem in mobile phone based
tokens by having multiple client keys, both in parallel and over time
(Req.~\ref{req:identity-binding}). This leads to a requirement that PIN policy
and PIN handling are linked to keys only by association, e.g.\ using PIN values
that are shared among many keys and changeable independently from the key
objects they are associated with (Req.~\ref{req:one-to-many}).

\subsection{Security requirements}

Security requirements include minimum token key strength~\cite{Lenstra99}, and
the use of modern hash algorithms in identity certificates. For most
jurisdictions the minimum eID requirement today is equivalent to 2048 bit RSA
and SHA256 for the certificate signature. In this paper, we do not address these
requirements directly; for our purposes we assume that the underlying TPM 2.0
implementation supports the necessary primitives.

The requirements on the token platform have been summarized by Dimitrienko et
al.~\cite{Dmitrienko12}. These are the confidentiality of the identity key and
the code isolation for the identity integrity verification algorithm operating
on the identity key. The isolation must hold against software in the device, but
also with regards to other credentials that may be stored and used in the same
token. The access control to the credential, with respect to unauthorized users
as well as unauthorized code (e.g.\ malware) in the device must be guaranteed.
Last, the integrity of the credential code must be ascertained. Often it is also
necessary to ascertain the origin or integrity of the calling code as well.

Certification servers are subject to stringent security evaluation and
processes, and certificate enrolment is also rightly considered an important
security function, since it cements the proof of the binding between the
physical person and his identity token. Reliable attestation is therefore
paramount to proper enrolment (see Section~\ref{sec:attestation}).

\subsection{Privacy requirements}

Privacy requirements for eID~\cite{Naumann08} include \textit{preventing
unauthorized access} to the identity token and applying \textit{linkability
control} either in the eID client or at a server. Mechanisms to achieve the
latter includes providing identity \textit{assertions} rather than full identity
information as part of eID transactions, and, where applicable, keeping digital
citizen identity numbers separate from conventional national identification
numbers, e.g.\ social security numbers. One way to achieve this is the use of
service-provider specific asymmetric keys, as in the FIDO protocols. Protocol
confidentiality, i.e.\ using the identity only in the context of a secure
channel, safeguards against eavesdropping.

\section{Design}
\label{sec:design}

\subsection{Architecture}

\begin{figure*}
  \centering
  \begin{subfigure}{.5\textwidth}
    \centering
    \includegraphics[width=.95\linewidth]{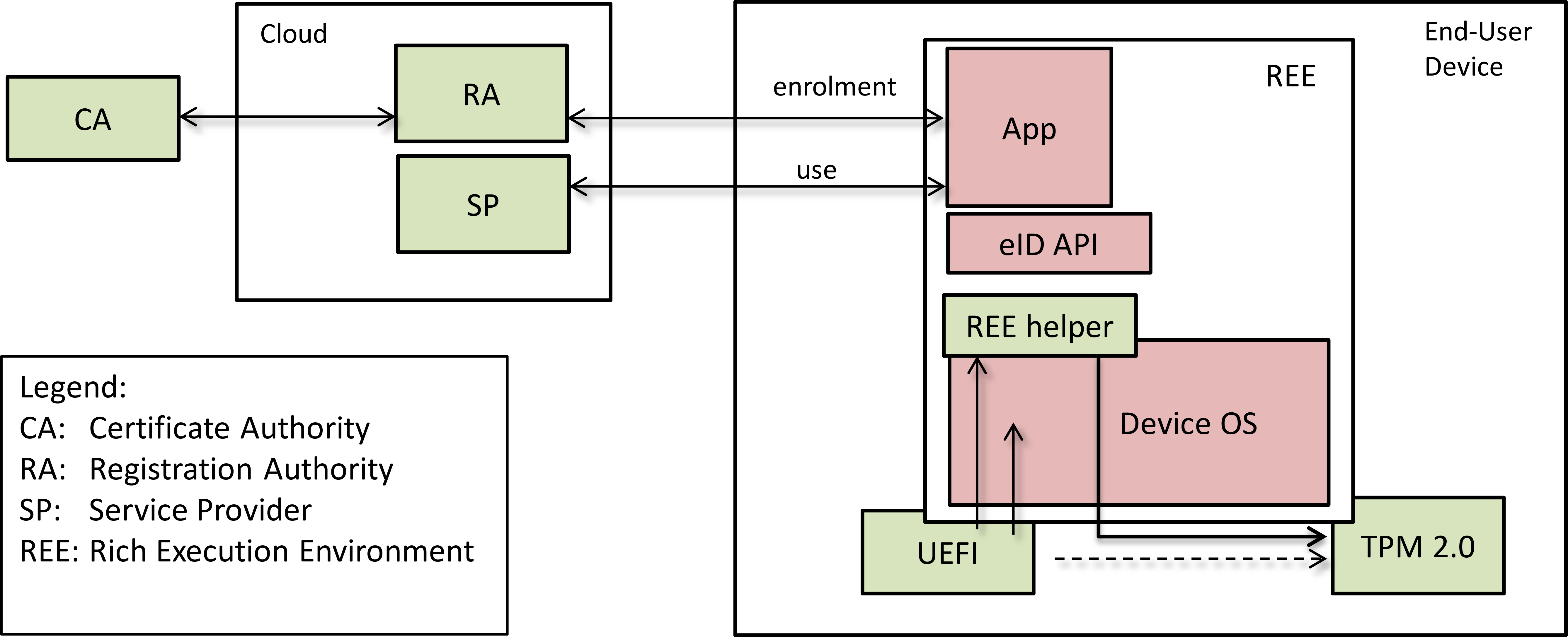}
    \caption{System architecture}
    \label{fig:system-architecture}
  \end{subfigure}%
  \begin{subfigure}{.5\textwidth}
    \centering
    \includegraphics[width=.95\linewidth]{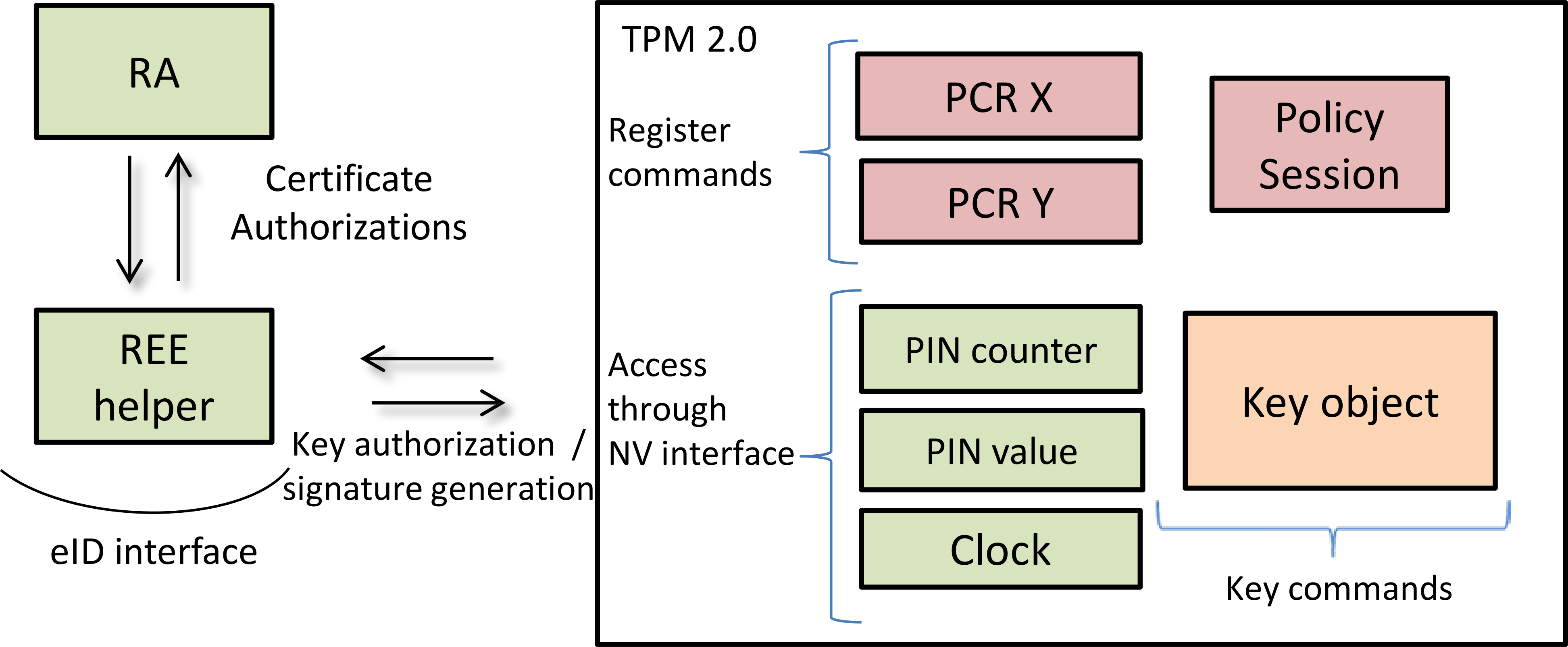}
    \caption{Device internal architecture}
    \label{fig:policy-components}
  \end{subfigure}
  \caption{eID architecture}
\end{figure*}

Our design addresses the security demands for an eID token, as outlined in
Section~\ref{sec:requirements}. We operate in an system architecture illustrated
in Figure~\ref{fig:system-architecture}. The device contains a TPM 2.0.
According to specification, this module operates in isolation from the (REE)
operating system, and has its own non-volatile storage, random number source and
needed cryptographic primitives. The implementation of the TPM 2.0 module can be
separate hardware, or, especially in mobile devices, it can be implemented
inside a TEE\@. For the purposes of this paper, the chosen implementation path
is irrelevant.

According to TPM specifications, the end-user device must provide a Core Root of
Trust for Measurement (CRTM) that supplies basic platform software integrity
measurements to the TPM in a reliable manner. This function is provided for the
TPM by the platform's authenticated or secure boot-up code, and is in
Figure~\ref{fig:system-architecture} represented by the \emph{Unified Extensible
Firmware Interface} (UEFI) secure boot~\cite{Wilkins13}. For the purpose of the
eID example, reliable measurements of \textbf{a)} the OS and \textbf{b)} a
component we call \emph{REE helper} need to be provided to the TPM\@. The
version of the OS (and e.g.\ its TPM drivers) and the REE helper that will
run our TPM policy represents the part of the software stack that serves as a
precondition for the TPM 2.0-based eID system to work. We expect an
authorization of the fact that necessary components are in place and that e.g.\
user PINs can be entered by the user at some level of assurance. We will see
that the integrity of the REE components are not otherwise a requirement
for the policy or the eID function to work correctly.

Our eID system also contains off-device components. The \emph{Registration
Authority} (RA) is the network facing entity that provides credential enrolment.
As part of the enrolment process the RA also performs attestation of the device
endpoint. As was highlighted in Section~\ref{sec:attestation}, the details of
platform attestation are out of scope for this paper --- we assume that the TPM
has an EK for which an associated platform certificate exists.  Through this
certificate, the RA can, with standard TPM 2.0 functionality, assert the
platform dependence of the enrolled eID authentication keys. In our architecture
we also associate a number of remote authorizations for the TPM EA policy as
being produced by the RA\@. Some of these authorizations are permanent, others
need to be dynamically refreshed. In a production system this off-device Trusted
Third Party (TTP) that produces the authorizations can be distributed, and need
not necessarily always be associated with the RA function, but it nevertheless
is an integral part of the security of our design.

The \emph{Certification Authority} (CA) is the entity that provides the user
certificate for the eID key as part of the enrolment process. It is common
practice that the CA is not combined with the RA, and often the CA is physically
isolated from the network used by the other entities in the system.

The \emph{Service Provider} (SP) is the entity that leverages the eID, by
validating the certificate and a signature by the eID key on a fresh challenge
provided by the SP\@. A very typical use case in this context is a
client-authenticated TLS sessions to a government service or a
bank\footnotemark.

\footnotetext{eID is also often used for document signatures, in which case the
role of the SP becomes slighly different.}

The REE helper is an important component in our design, although not many
security requirements are put on it. Since the TPM interfaces, and especially
policy generation are quite different from what can be expected by ordinary
developers mostly interested in making his/her application eID-aware, the REE
helper provides the interface typically used for ID access and use --- the
PKCS\#11 standard, and operates to fulfill the requirements identified in
Section~\ref{sec:requirements}. The principal function of the REE helper is
therefore to translate the eID APIs and requirements into TPM library commands
and concepts. Since the usage policy in our example will be mapped to TPM EA
policy, we do not have to require perfect integrity of the REE helper. In case
its integrity is compromised, the main possible consequence is that the TPM
policies will not be satisfied, leading to a \emph{Denial of Service} (DoS)
situation.

\subsection{Authorization Policy}
\label{sec:authorization-policy}

We discussed the basics of the TPM 2.0 EA policies in
Section~\ref{sec:tpm2-extended-authorization}. In this section we will describe
how to to express the following aspects of our requirements using EA policy
primitives:

\begin{itemize}
  \item Platform integrity \hfill (Section~\ref{sec:platform-integrity})
  \item Restricting the number of PIN attempts \hfill (Section~\ref{sec:pin-attempts})
  \item Reset of the PIN entry counter \hfill (Section~\ref{sec:pin-counter-reset})
  \item PIN entry and comparison \hfill (Section~\ref{sec:pin-entry})
  \item Credential revocation (expiration) \hfill (Section~\ref{sec:credential-revocation})
\end{itemize}

Figure~\ref{fig:policy-components} illustrates the TPM parameters and objects we
will use in the following description. Each following section is accompanied by
a listing of the TPM commands invoked by the REE-helper to satisfy the policy.
In these listings, each command is represented in the form:

\vspace{\baselineskip}
\noindent
\resizebox{\linewidth}{!}{%
  $[digest\;\leftarrow]\;command(<in\;parameters>[,\;<out\;parameters>])$
}

\subsubsection{Platform integrity}
\label{sec:platform-integrity}

\begin{lstlisting}[float=tb,label={lst:platform-integrity},
caption={Platform integrity policy fragment}]
@$d_{0}\leftarrow          $@TPM2_StartAuthSession(
@\hphantom{$d_{x}\leftarrow$}@  <@$keyHandle_{eID}$@,@$sessionType_{policy}$@>,)
@\hphantom{$d_{x}\leftarrow$}@  <@$sessionHandle_{eID}$@>)
@$d_{1}\leftarrow          $@TPM2_PolicyPCR(<@$sessionHandle_{eID}$@, @$pcrSelection$@>)
@\hphantom{$d_{x}\leftarrow$}@[@$\ldots$@]
@\hphantom{$d_{x}\leftarrow$}@TPM2_LoadExternal(<@$pubKey_{A}$@>, <@$keyHandle_{A}$@>)
@\hphantom{$d_{x}\leftarrow$}@TPM2_VerifySignature(
@\hphantom{$d_{x}\leftarrow$}@  <@$keyHandle_{A}$@, @$sig_{a}$@>, <@$ticket_{\{a,A\}}$@>)
@$d_{2}\leftarrow           $@TPM2_PolicyAuthorize(
@\hphantom{$d_{x}\leftarrow$}@  <@$sessionHandle_{eID}$@, @$signature_{a}$@, @$policyRef_{a}$@,
@\hphantom{$d_{x}\leftarrow$}@   @$keyName_{A}$@, @$ticket_{\{a,A\}}$@>)
\end{lstlisting}

The baseline of the policy (see Listing~\ref{lst:platform-integrity}) is the
invocation of one or more \texttt{TPM2\_PolicyPCR()} commands. The PCR registers
are reliably populated by boot-up measurements by the CRTM, and these PCR values
cannot be re-populated to any known values after boot. Therefore, a
\texttt{TPM2\_PolicyPCR} command that maps the values of these PCRs into the
policy session will identify whether the device was booted into a software state
recognizable as part of the policy, or into some other state.

To make the policy modular, e.g.\ to make it possible to update the REE helper
to a newer version, it makes sense to partition the policy collection with
external authorizations. In Listing~\ref{lst:platform-integrity}, this is based
on an external public key which is loaded to the TPM using the
\texttt{TPM2\_LoadExternal()} command. Now, the external authorization can occur
as described in Section~\ref{sec:external-authorization}; first, the signature
of the external authorization, which is of a standardized format, is validated
with \texttt{TPM2\_VerifySignature()}; second the locally signed ticket produced
by \texttt{TPM2\_VerifySignature()} is passed to
\texttt{TPM2\_PolicyAuthorize()}.  As we can recall from our earlier example, if
the validated authorization matched the current \texttt{policyDigest} for the
session, the \texttt{policyDigest} is reset, and replaced, essentially, by the
public key hash of the public key used to verify the signature.

\subsubsection{Restricting the number of PIN attempts}
\label{sec:pin-attempts}

\begin{lstlisting}[float=tb,label={lst:pin-attempts},
caption={PIN attempt restriction policy fragment}]

@$d'_{0}\leftarrow          $@TPM2_StartAuthSession(
@\hphantom{$d_{x}\leftarrow$}@  <@$objectHandle_{ctr}$@,@$sessionType_{policy}$@>, <@$sessionhandle_{ctr}$@>)
@$d'_{1}\leftarrow          $@TPM2_PolicyCommandCode(<TPM_CC_NV_READ>)
@\hphantom{$d_{x}\leftarrow$}@TPM2_NV_Read(<@$sessionhandle_{ctr}$@, @$ctr$@>, <@$n$@>)
@\hphantom{$d_{x}\leftarrow$}@
@$d_{3}\leftarrow           $@TPM2_PolicyNV(<@$sessionHandle_{eID}$@, @$ctr$@, @$n$@, @$eq$@>)
@\hphantom{$d_{x}\leftarrow$}@
@$d''_{0}\leftarrow          $@TPM2_StartAuthSession(
@\hphantom{$d_{x}\leftarrow$}@  <@$keyHandle_{ctr'}$@,@$sessionType_{policy}$@>, <@$sessionHandle_{ctr'}$@>)
@$d''_{1}\leftarrow         $@TPM2_PolicyCommandCode(<TPM_CC_NV_Increment>)
@\hphantom{$d_{x}\leftarrow$}@TPM2_NV_Increment(<@$sessionhandle_{ctr}$@, @$ctr$@>)
@\hphantom{$d_{x}\leftarrow$}@
@$d_{\underset{(n,n+1)}{}}$\hspace{-.68cm}$\leftarrow$@TPM2_PolicyNV(@$sessionHandle_{eID}$@, <@$ctr$@, @$n+1$@, @$eq$@>)
@$d_{4}\leftarrow           $@TPM2_PolicyOR(<@$sessionHandle_{eID}$@, @$d_{(0,1)}$@,@$d_{(1,2)}$@,@$d_{(2,3)}$@>)
@\hphantom{$d_{x}\leftarrow$}@[@\textit{counter reset} (see Section~\ref{sec:pin-counter-reset})@]
@\hphantom{$d_{x}\leftarrow$}@TPM2_VerifySignature(
@\hphantom{$d_{x}\leftarrow$}@  <@$keyHandle_{A}$@, @$signature_{b}$@>, <@$ticket_{\{b,A\}}$@>)
@$d_{5}\leftarrow           $@TPM2_PolicyAuthorize(
@\hphantom{$d_{x}\leftarrow$}@  <@$sessionHandle_{eID}$@, @$signature_{b}$@, @$policyRef_{b}$@,
@\hphantom{$d_{x}\leftarrow$}@   @$keyName_{A}$@, @$ticket_{\{b,A\}}$@>)
\end{lstlisting}

The next policy aspect we consider is the number of failed PIN attempts for the
key we intend to use (Req~\ref{req:pin-and-puk}). To capture this we use a NV
storage element configured to act as a counter, i.e.\ during its initialization
it is associated with an \texttt{TPMA\_NV\_COUNTER} attribute. This makes writes
to the NV element only possible in a monotonically increasing manner using the
\texttt{TPM2\_NV\_Increment()} command. We also assume that the NV object
holding the counter is associated with an authorization policy of its own, where
resetting the counter is conditional to the entering of the correct PIN value as
part of the policy, but incrementing the counter is allowed for the REE helper
without restriction. The specifics of this policy are described in
Section~\ref{sec:pin-counter-reset}. For now, the goal is to bind proof of
increment (that the counter has been incremented by one during the policy
session) into the \texttt{policyDigest}, and assert that the final counter value
is less than, or equal to the allowed number of PIN attempts (see
Listing~\ref{lst:pin-attempts}). Note that the invocation of
\texttt{TPM2\_NV\_Increment()} cannot be directly determined in the policy
session for the eID credential. Hence a mere comparison between the current
counter value and the allowed number of attempts is insufficient to ensure
proper operation. Instead, we use \texttt{TPM2\_NV\_Read()} to retrieve the
current counter value in the REE helper. Then we bind the current counter value
to the policy digest of the main policy session using \texttt{TPM2\_PolicyNV()},
increment the counter in a separate session, then proceed to bind the now
incremented counter value to the main session policy digest. Recall from
Section~\ref{sec:tpm2-extended-authorization} that \texttt{TPM2\_PolicyNV()}
makes its comparison immediately, not in a deferred manner.

Let us assume that the maximum number of (failed) PIN entry attempts allowed is
three. In this case, the previous policy segment should accept combinations of
the NV storage element value pairs (0,1), (1,2) and (2,3).  All of these
represent different acceptable \texttt{policyDigest} values for the policy
session. The values can be collapsed into one using the
\texttt{TPM2\_PolicyOR()} conjunction (see Section~\ref{sec:policy-or}). The REE
helper, possibly with the aid of the RA, can compute these expected values,
which each is the result of the consecutive application of the prior
\texttt{TPM2\_PolicyAuthorize()} and two \texttt{TPM2\_PolicyNV()} invocations
for one of the above pairs. The resulting \texttt{policyDigest} value is now
independent of the value pair that was encoded into the \texttt{policyDigest}
prior to the \texttt{TPM2\_PolicyOR()}.

For consistency, we again normalize the policy construction by the application
of another external authorization originating from the RA\@. The reasoning for
this is similar to the first stage normalization -- for the PIN count we can
envision variations in e.g.\ which NV storage element is used as the PIN
counter, or possibly the maximum allowed attempts. Note that signature used for
this normalization should include a different \texttt{policyRef} from the one
applied before. This is needed so that the TPM implicitly protects against
policy fragment re-ordering or the omission of some fragments in the middle of a
policy session.

\subsubsection{PIN counter reset}
\label{sec:pin-counter-reset}

On each attempted use of the eID signature key, the PIN entry counter is
incremented. If the use of the signature key succeeded, we know that \textbf{a)}
the PIN received at the REE helper was correct, and \textbf{b)} that the counter
value had not passed the threshold for allowed tries. If the threshold was
exceeded, we require that a PUK value is entered correctly for the PIN counter
to be reset. For this policy, we formalize the PUK as high-entropy
system-generated password value that requires no inherent replay protection.
However, as shown by the PIN handling in the main policy, the PUK could equally
well be modelled like the PIN, with the same replay-protection setup.

The TPM 2.0 specification does not provide the possibility to reset a counter NV
object, i.e.\ the actual reset operation must be split into two parts; the
erasure of the counter object (to which an associated policy is applied), and
its recreation. The object recreation is conditioned to ``platform-specific''
user authorization, which may be vary from completely open to any
platform-issued policy, i.e.\ as a worst case we must assume it to be completely
unprotected. Herein lies one threat against our system for PIN replay
protection. As the counter reset is not atomic, if an attacker can disrupt the
object creation e.g.\ power-cycle the device with a physical reset switch at
exactly the right moment after a successful authentication, the attacker might
be able to recreate the counter with a policy that allows reset with less
privileges. This sets the stage for brute-force PIN resolving, which in turn
may lead to unauthorized use of the signing key. This also constitutes one
reason for why the integrity measurement of the REE helper (see
Section~\ref{sec:platform-integrity}) matters in the overall policy. Considering
the complexity of the attack, we argue that an attacker of this strength may do
better by just eavesdropping the PIN value when entered by the user. There are
also processes for command audits in the TPM 2.0 specification that may be
applied as protection for the atomicity problem, but these require another
internal or external trusted environment for validation.

The initial phase of the policy construction for PIN counter reset mirrors that
of key usage. First, the integrity of the system and the REE helper is assured,
followed by the external authorization, as outlined in
Section~\ref{sec:platform-integrity}. The second phase has two alternative
paths. The first one mirrors the updating of the PIN retry counts but sets the
upper limit to $n+1$, i.e.\ if the limit was 3, as in the earlier example, we
use 4 as the limit here. This is followed by the entering of the PIN, described
in \texttt{Section~\ref{sec:pin-entry}}. The second path is the PUK entry
option. Here, the initial value composition also include the integrity checks
for the system and REE helper. This is followed by binding the preferably
strong PUK password to the policy by the application of a
\texttt{TPM2\_PolicySecret()} in a manner equivalent to PIN entry (see
Section~\ref{sec:pin-entry}.

The two possible policySession values are then combined using
\texttt{TPM2\_PolicyOR()}, possibly followed by a
\texttt{TPM2\_PolicyAuthorize()} for good measure. The final step of the policy
will have to account for the current state of the session augmented with
\texttt{TPM2\_PolicyCommandCode()} for the command
\texttt{TPM2\_NV\_UndefineSpace}. The resulting session value must be \textbf{OR}ed with
the completely open policy for counter updates and reads, i.e.\ a policy with
\texttt{TPM2\_PolicyCommandCode()} for the commands \texttt{TPM2\_NV\_Increment}
or \texttt{TPM2\_NV\_Read}. Finishing up the object \texttt{authPolicy} value
after this follows the main policy for key use.

PIN updates also follow from the right to remove the NV object holding the
PIN\@. As defined in this case, those objects have secret-based authorization,
i.e.\ erasing them requires only the knowledge of the respective PIN or PUK\@.
The recreation of the NV record, i.e.\ adding a new PIN with
\texttt{TPM2\_NV\_DefineSpace()} may require user authorization (a user
password).

\subsubsection{PIN entry}
\label{sec:pin-entry}

The next stage of the policy after PIN counter handling is the entry of the
PIN\@. There are two dedicated policy commands for this, the
\texttt{TPM2\_PolicyAuthValue()} and \texttt{TPM2\_PolicyPassword()} both of
which, with slight variations, defer a mandatory check of the PIN for the
authorized key object before object use (recall that most TPM 2.0 objects can
contain a password value for authorization, this is a feature that is inherited
from the TPM 1.2 specification). However, associating the PIN value with the key
object is not acceptable, as we want to separate PIN objects from keys to allow
a particular PIN to be shared by many keys (Req.~\ref{req:one-to-many}, and
manage the life cycle of these PINs (e.g.\ change of the PIN value) in isolation
from the eID key objects referring to them. The TPM 2.0 specification provides a
policy command ideally suited for this kind of setup:
\texttt{TPM2\_PolicySecret()} lets us associate the knowledge of a secret,
encoded in another object's \texttt{authValue} with the policy of the object we
are intending to use. The policy digest will be updated with the name of the
object we associating to. In principle we could associate to any kind of object,
e.g.\ a ``dummy'' key, but using a NV record for this purpose has the advantage
that its name does not change when re-created, whereas the public key component
of an asymmetric key pair is represented in its name.  Thus we use an empty NV
record with the PIN as its \texttt{authValue} as the associated object. On
correct PIN entry (see Listing~\ref{lst:pin-entry}) the REE helper is empowered
with the knowledge necessary to both validate, erase and replace the NV object
with the PIN, and it is up to the integrity protection of the REE helper to
ascertain its correct operation with regards to using (and erasing the PIN from
its memory) in the different TPM operations involving PINs and
PUKs\footnotemark. This part of the policy may once again be rounded off with
the application of \texttt{TPM2\_PolicyAuthorize()}, to potentially allow e.g.\ for device-specific NV index
variations for PIN handling.

\begin{lstlisting}[float=tb,label={lst:pin-entry},
caption={PIN entry policy fragment}]
@$d'''_{0}\leftarrow        $@TPM2_StartAuthSession(
@\hphantom{$d_{x}\leftarrow$}@  <@$objectHandle_{PIN}$@, @$sessionType_{policy}$@>,
@\hphantom{$d_{x}\leftarrow$}@  <@$sessionHandle_{PIN}$@>)
@$d'''_{1}\leftarrow        $@TPM2_PolicyPassword(
@\hphantom{$d_{x}\leftarrow$}@  <@$sessionHandle_{PIN}$@, @$authValue_{PIN}$@>)
@\hphantom{$d_{x}\leftarrow$}@
@$d_{6}\leftarrow           $@TPM2_PolicySecret(
@\hphantom{$d_{x}\leftarrow$}@  <@$sessionHandle_{eID}$@, @$objectHandle_{PIN}$@,
@\hphantom{$d_{x}\leftarrow$}@   @$sessionHandle_{PIN}$@>)
@\hphantom{$d_{x}\leftarrow$}@TPM2_VerifySignature(
@\hphantom{$d_{x}\leftarrow$}@  <@$keyHandle_{A}$@, @$signature_{c}$@>, <@$ticket_{\{c,A\}}$@>)
@$d_{7}\leftarrow           $@TPM2_PolicyAuthorize(
@\hphantom{$d_{x}\leftarrow$}@  <@$sessionHandle_{eID}$@, @$signature_{c}$@, @$policyRef_{c}$@,
@\hphantom{$d_{x}\leftarrow$}@   @$keyName_{A}$@, @$ticket_{\{c,A\}}$@>)
@$d_{8}\leftarrow           $@TPM2_PolicyCommandCode(<TPM_CC_Sign>)
@\hphantom{$d_{x}\leftarrow$}@TPM2_Sign(<keyHandle, digest>, <signature>)
\end{lstlisting}

\footnotetext{The obvious vulnerabilities with PIN handling in the REE can be
remedied by hardware-supported secure PIN entry. This ``Secure UI'' can be
mapped to the TPM concept of a higher \textit{locality} -- a binding that can be
included in TPM EA policy. Such setups are for now highly device-specific, and
we do no consider this further in this paper.}

Finally, \texttt{TPM2\_PolicyCommandCode()} (for \texttt{TPM2\_Sign()}) is
invoked to limit the capabilities the authorization sessions grants on the key
object. The final session policy value (which is an aggregation of the
device-specific license authorization (public key) and the command code
constraint) is the value that shall be added to the key object when the key is
created by \texttt{TPM2\_Create()}. The value can be attested by the RA, since
key creation can happen in a remote-originated authenticated channel, and the
value can be confirmed by returned attributes (e.g.\ \texttt{creationData}). In
fact, if the RA makes sure that no two licenses issued to a specific endpoint
are alike, all keys can be created with the same \texttt{authPolicy} value. The
policy construct is flexible enough to account for all optionality as part of
the individual RA-originating authorizations. This even holds when the policy is
augmented with extra functionality, e.g.\ such as described in the following
section.

\subsubsection{Credential revocation}
\label{sec:credential-revocation}

With removable tokens such as smart cards, credential revocation often relies
solely on certificate revocation, and elaborate protocols and constructs such as
\emph{Certificate Revocation Lists} (CRLs) and on-line protocols such as the
\emph{On-Line Certificate Status Protocol} (OCSP) allow for service providers to
confirm the revocation status of a credential within the lifetime indicated in
its certificate. However, with TEEs and TPMs, credential revocation can also be
managed at the client-side as an additional security measure. The TPM 2.0
specification has a millisecond-resolution clock that is guaranteed to advance
when the TPM is powered on, but it also has periodic backup to NV memory, i.e.\
within certain tolerance it behaves like a secure clock. Policy access to this
clock is through the \texttt{TPM2\_PolicyCounterTimer()} policy command. The
command behaves as a timer, i.e.\ a time comparison is made based on ``time
passed'' since a reference value which is provided in the command parameters,
and which also is reflected in the policy value update. This can be
straight-forwardly mapped to a subsequent \texttt{TPM2\_PolicyAuthorize()} with
a device-specific key originating from the RA, i.e.\ the RA can this way provide
time-dependent licenses for the usage of the eID key that are more short-lived
than the CA-generated certificate --- a ``dead man's switch'' of sorts. This
approach makes practical sense, since it eases the interactive load on service
providers, who now can trust that signatures against eID certificates residing
in TPMs (a fact that can be logged in an certificate attibute extension) have
this ``automatic'' revocation property, i.e.\ the keys stop being usable if key
revocation occurs within the time period of the CA-issued certificate.

\section{Analysis}
\label{sec:analysis}

We limit the scope of our analysis to the presented eID requirements in the
context of constructs described in Section~\ref{sec:authorization-policy}.
While we acknowledge that formal verification of the TPM 2.0 EA primitives is a
worthy pursuit, we consider it out of scope for this paper. Likewise, as all
TPM 2.0 platforms are by definition required to provide platform assurance, it
is not in our power to add to or detract from the trustworthiness of the
platform itself.

In terms of functional requirements, the EA policy binds key authorization to
any number of shared or disjunct PINs and PUKs, and as the TPM by internal
design can support any number of key objects, we claim that requirements by both
EU eID and FIDO are easily met. Necessary security guarantees are provided by
the TPM 2.0 itself. Of the privacy requirements, we do address linkability during
token use. There may also be linkability concerns during enrolment and platform
authentication, these are issues external to this paper, but TPM 2.0 does provide
optional \emph{Direct Anonymous Attestation}~\cite{Chen13} (DAA) support to
mitigate even this threat. For credential user binding, we leverage TPM 2.0 EA,
and provide the example as proof that we can provide retry-protection akin to
that of PKCS\#15 tokens.

Furthermore, we provide the dead man's switch - type behaviour for the eID
credential. This partially offline system is one way of allowing eID to be
reliably used with SPs that have only point-to-point connectivity with the ESP,
such as physical doors or out-door event ticket validators.

We also want to highlight the relaxed integrity requirements on the REE helper
as an important side-effect of leveraging TPM 2.0 EA\@. Even as the presented
authorization model involves a large number of commands and operations, the
entities securing the authorization are the policy sessions, in turn proteced by
TPM 2.0. Except for the atomicity issue highlighted in Section~\ref{sec:design},
all operational misbehaviour from the side of the REE helper constitutes only
Denial-of-Service.

\section{Perceived EA shortcomings}
\label{sec:perceived-ea-shortcomings}

Despite the flexibility of the TPM 2.0 EA mechanism, on several occasions we
find the need to resort to awkward constructs in order to formulate primitives
needed for our policy. We identify the following instances where alternative
design decisions than the ones made by the TPM 2.0 designers could have made
policy constructions simpler and more intuitive:

\noindent\textbf{Including a secret value as part of a policy} The current TPM
2.0 specification EA policy command set does not allow a secret value to be
included as part of the policy. A password entry, for instance, could be
trivially turned into a policy command, hashing that password directly into the
session policy value. Such a command could have found immediate use in our
design, allowing us to avoid the use of ``dummy'' NV objects in
Section~\ref{sec:pin-entry} to hold the PIN and PUK.

\noindent\textbf{Linking two consecutive TPM commands} The current specification
does not allow a TPM command to be conditional on the outcome of another.  For
example, with such a feature, we could have specified a policy: ``reset PIN
counter \emph{only} after successful sign-operation'' which could have made the
PIN counter reset detailed in Section~\ref{sec:pin-counter-reset} simpler, and
thus stronger.

\noindent\textbf{Access rights for comparing TPM objects} Another shortcoming
that we encountered, is that comparisons between TPM objects, especially NV
elements, can only be done while obtaining read rights to the object being
referred to. This makes sense from a data-flow standpoint, but makes many policy
constructs cumbersome or even impossible.

\section{Related work}
\label{sec:related-work}

The legal, technical and organizational challenges in implementing a pan-European
eID framework have been analyzed in a number of studies. The study by
Myhr~\cite{Myhr08} mainly focuses on the issuance procedures and the lack of a
common unique identifier for physical persons on an European level that could be
used in eID. Since the publication of the study, a large study was conducted by
the European Union regarding eID interoperability for \emph{Pan-European
E-Government Services} (PEGS)\footnotemark. The results of this study are a
final report consisting of analysis and assessment of eID interoperability
requirements and 32 distinct country profiles on national schemes from both a
legal and technical perspective. Mahler~\cite{Mahler13} in turn proposes a
multi-stakeholder governance model for an European eID influenced by
multi-stakeholder institutions used in Internet governance, such as the ICANN
and IETF\@.

\footnotetext{\url{http://ec.europa.eu/idabc/en/document/6484.html}}

There are also a few technical studies with the aim of overcoming some of the
limitations and challenges with conventional, smart card-based identity tokens.
\emph{TPMident}~\cite{Klenk09} is a two-factor authentication system based on
TPMs, with the goal of protecting users against identity theft. In TPMident,
conventional eID is used to establish initial trust to a \emph{non-migratable}
authentication credential stored within a TPM\@. The authors have also
integrated TPMident with the \emph{OpenID}\footnotemark\ sign-on protocol.

Dmitrienko et al.~\cite{Dmitrienko12} present the design of a token-based access
control system for NFC-enabled smartphones which allows users to maintain access
control credentials for multiple resources. A key feature of their scheme is the
ability delegate access rights to other smartphone users without the involvement
of a central authority (RA).

Vossaert et al.~\cite{Vossaert14} propose a system for secure PIN-entry for
smart cards. The solution relies on a workstation equipped with a TEE and an
attached smart card reader. The authors present a proof-of-concept prototype of
the system using Belgian eID cards and PCs with TPM-based secure execution
environments. 

\footnotetext{\url{http://openid.net/}}

\section{Conclusions}
\label{sec:conclusion}

On one hand, our eID architecture demonstrates that the new TPM 2.0 Enhanced
Authorization model can meet the requirements of a widely deployed, real-world
use-case. On the other, despite the richness of the current EA model, we
identify some possible improvements that would enable simpler, and thus more
secure, solutions. We believe that adding TPM 2.0 to the set of possible ESP
implementations will add a large class of (mobile) devices as eID endpoints.
  
In Section~\ref{sec:analysis} we identified some shortcomings of the EA model
that, if addressed in TPM specifications, can increase the flexibility of TPM 2.0
EA even further. Even with such improvements, we still have to recognize the
difference between full token programmability in an isolated, secure environment
and a configurable authorization model to a well defined and analyzed set of
security primitives. Both approaches have their individual merits on the axes of
flexibility vs.\ security.

As future work, we plan to complete a working Proof-of-Concept ESP as presented
in this paper, to be trialed against a government PKI\@. Other research avenues
include integrating biometrics and other trustworthy I/O to the ESP design.

\section{Acknowledgments}
\label{sec:acknowledgements}

This work was financially supported in part by the Intel Collaborative Research
Institute for Secure Computing and by the Academy of Finland, project No 283135
(CloSe: Cloud Security Services).

\ifx\istechreport\undefined
An extended version of this paper is available as a research
report~\cite{Nyman14}.
\fi

\bibliographystyle{hacm}
\bibliography{eid}

\ifdefined\istechreport
\appendix

\section{TPM 1.2 Authorization Protocols}
\label{appendix:tpm-authprotocols}

The TPM 1.2 specification includes three protocols to securely pass
proof-of-knowledge of the owner authorization value; the
\emph{Object-Independent Authorization Protocol} (OIAP), the
\emph{Object-Specific Authorization Protocol} (OSAP), and the
\emph{Delegate-Specific Authorization Protocol} (DSAP). 

\subsection*{OIAP}

The OIAP protocol supports multiple authorizations sessions to arbitrary
entities; in other words, a single OSAP session can be used to invoke multiple
commands on the TPM, which manipulate different objects. However, an OSAP
session cannot be used to introduce new authorization information to the TPM
 (e.g.\ key creation), as the owner authorization value of the entity being
accessed is used to key the Hash-based Message Authentication Code (HMAC) used
for authentication and integrity verification of session traffic.

\subsection*{OSAP}

The OSAP protocol supports authorization sessions for a single entity, i.e.\ all
commands invoked during a particular OSAP session may only manipulate the same
entity. The advantage is that OSAP enables the confidential transmission of new
authorization information. OSAP session initialization involves the creation of
a ephemeral secret (henceforth referred to as the OSAP shared secret) used to
protect further protocol traffic. The purpose of this arrangement is two-fold;
first, it allows the caller to cache the OSAP shared secret for the duration of
an extended session without compromising the owner authorization value of the
entity being accessed; second, it enables new authorization information to be
inserted into the TPM\@. The insertion of new authorization data is covered by
the \emph{AuthData Insertion Protocol} (ADIP), which involves encryption of the
new authorization data using a one-time key generated computing a hash of the
OSAP session secret and a session nonce. A similar scheme is utilized by the
\emph{AuthData Change Protocol} (ADCP) to change the owner authorization
associated with an existing entity. Knowledge of the old owner authorization
value is required as a proof-of-ownership in order to change the authorization
information of an entity. As reusing the OSAP shared secret multiple times could
expose it to cryptoanalytic attacks, the TPM is responsible of terminating OSAP
sessions that have introduced new authentication data with ADIP or ADCP\@. The
caller is responsible for terminating any authorization session to ensure the
removal of any saved authorization information and ephemeral secrets.

\subsection*{DSAP}

The DSAP protocol allows access to an entity to be delegated without disclosing
the owner authorization value to a third party. The owner of the key may dictate
the set of commands that can be performed using the delegated cryptographic key
when authorization other than actual owner authorization is used. For instance,
the recipient of a delegated key may be allowed to use the key for encryption or
decryption, but may not delegate the key further or modify the authorization
information associated with the key. This is realized via the creation of a
\emph{delegation structure} by the key owner. The delegation structure specifies
the cryptographic key to be delegated, the set of commands that may be performed
using the key, and a \emph{delegation authorization value}. The delegation
structure itself is encrypted with a key known only to the TPM\@. The purpose of
this is to protect the confidentiality of the delegation authorization value
associated with the delegation structure, as any caller who can supply the
structure, and has knowledge of the associated delegation authorization value,
may use the delegation structure to perform the operations specified by the
structure on the delegated key. This allows the delegation mechanism to operate
without changes to TPM commands which expect a authorization value to be
specified upon invocation.

Similarly to OSAP session, DSAP protocol session are restricted to a single
entity. During DSAP session initialization, the caller specifies the target
entity, and supplies the delegation structure for that object. Also similarly to
OSAP, the DSAP session initialization involves the creation of an ephemeral
secret used to protect further traffic. During the DSAP authorization session,
the caller may invoke any commands allowed by the specified delegation structure
by supplying the delegation authorization value of that structure \emph{instead
of} the owner authorization value associated with the object being accessed.

\section{TPM 2.0 Extended Authorization Commands}
\label{appendix:extended-authorization-commands}

This section provides a brief overview of the Extended Authorization commands
referred to elsewhere in this paper.

\subsection{\texttt{TPM2\_PolicyAuthorize()}}

The \texttt{TPM2\_PolicyAuthorize()} command authorizes changes to an existing
policy by an external entity. The authorizing entity signs the digest value it
wants to authorize. The TPM will verify the signature, and if valid, the
\texttt{TPM2\_PolicyAuthorize()} will reset the current \texttt{policyDigest}
value and update it with a digest derived from the (unique) name of the signing
key:

\vspace{.5\baselineskip}
\noindent
\resizebox{\linewidth}{!}{%
  $\begin{aligned}
    \mathbf{if}&(\mathbf{V}(keyHandle_{A}, signature, policySession \rightarrow policyDigest)\;\\
    &\;\;policyDigest_{new} = \mathbf{H}(0\;||\;TPM\_CC\_PolicyAuthorize\;||\;keyName_{A})\\
    &\;\;policyDigest_{new+1} =\mathbf{H}(policyDigest_{new}\;||\;policyRef)
  \end{aligned}$
}

\vspace{.5\baselineskip}
\noindent where

\begin{description}[itemsep=.1cm]
  \small
  \item[$TPM\_CC\_PolicyAuthorize$] \hfill is the policy command code
  \item[$A$]                 \hfill is the authorization key
  \item[$\mathbf{V}()$]      \hfill is the signature verification algorithm for $A$
  \item[$keyHandle_{A}$]     \hfill is the handle for $A$
  \item[$signature$]         \hfill is the signature of the policy to be authorized
  \item[$keyName_{A}$]       \hfill is the name of the object corresponding to $A$
  \item[$policyRef$]         \hfill is the policy classifier
\end{description}

External authorization is discussed in more detail in
Section~\ref{sec:external-authorization}.

\subsection{\texttt{TPM2\_PolicyAuthValue()}}

The \texttt{TPM2\_PolicyAuthValue()} command is a deferred assertion that sets
the \texttt{policySession} $\rightarrow$ \texttt{isAuthValueNeeded} property.
This indicates a caller provided \texttt{authValue} will be compared the
\texttt{authValue} of the authorized object when the policy session is used for
authorization. This property also indicates that the caller provided
\texttt{atuhValue} is included in the key used to calculate the authorization
HMAC for the current session. This key is provided as parameter to
\texttt{TPM2\_BeginAUthSession}. Once set, the \texttt{isAuthValueNeeded}
property may only be unset by \texttt{TPM2\_PolicyPassword()} or
\texttt{TPM2\_PolicyRestart()}, which returns the entire authorization session
to its initial state. The \texttt{policyDigest} derived by
\texttt{TPM2\_PolicyAuthValue} simple binds the corresponding command code as
part of the running digest:

\vspace{.5\baselineskip}
\noindent
\resizebox{\linewidth}{!}{%
  $policyDigest_{new} =
  \mathbf{H}(policyDigest_{old}\;||\;TPM\_CC\_PolicyAuthValue)$
}

\subsection{\texttt{TPM2\_PolicyCommandCode()}}

\texttt{TPM2\_PolicyCommandCode()} command is a deferred assertion that verifies
that the policy session is only used to authorize a particular command. The
command code for the authorized command is stored in \texttt{policySession}
$\rightarrow$ \texttt{commandCode} property. When the session is used for
authorization, the stored command code is compared to the code of the invoked
command. The stored command code is also bound to the \texttt{policyDigest}:

\vspace{.5\baselineskip}
\noindent
\resizebox{\linewidth}{!}{%
  $policyDigest_{new} =
  \mathbf{H}(policyDigest_{old}\;||\;TPM\_CC\_PolicyCommandCode\;||\;code)$
}

\vspace{.5\baselineskip}
\noindent where

\begin{description}
  \small
  \item[$TPM\_CC\_PolicyCommandCode$] \hfill is the policy command code
  \item[$code$]  \hfill is the code for command code being authorized
\end{description}


\subsection{\texttt{TPM2\_PolicyNV()}}

The \texttt{TPM2\_PolicyNV} is an immediate assertion, which performs an
arithmetic comparison between an input value and a value stored in a specified
non-volatile storage element. If the condition holds, the \texttt{policyDigest}
is updated accordingly:

\vspace{.5\baselineskip}
\noindent
\resizebox{\linewidth}{!}{%
    $\begin{aligned}
      \mathbf{if}&\;NValue\;\mathbf{op}\;operand\;\mathbf{then} \\
      &\;\;policyDigest_{new} = \mathbf{H}(policyDigest_{old}\;||\;TPM\_CC\_PolicyNV\;||
      \;args\;||\;NVName)
    \end{aligned}$
}

\vspace{.5\baselineskip}
\noindent where

\begin{description}
  \small
  \item[$TPM\_CC\_PolicyNV$]  \hfill is the policy command code
  \item[$args$]    \hfill is a hash over the input operands and operator
  \item[$NVValue$] \hfill is the value of the non-volatile storage element
  \item[$NVName$]  \hfill is the name of the non-volatile storage element
\end{description}

\subsection{\texttt{TPM2\_PolicyOR()}}

The \texttt{TPM2\_PolicyOR()} command authorizes one of several possible digest
values. It leads to an immediate assertion which verifies if the current
\texttt{policyDigest} value occurs in a list of authorized digest values
provided as input. If this is the case, the \texttt{policyDigest} is reset, and
updated to a value derived by calculating a digest of the concatenated input
digest values:

\vspace{.5\baselineskip}
\noindent
\resizebox{\linewidth}{!}{%
  $\begin{aligned}
    \mathbf{if}&\;policySession \rightarrow
    policyDigest\;\mathbf{in}\;digest_{1} \ldots digest_{n}\;\mathbf{then} \\
    &\;\;policyDigest_{new} =
    \mathbf{H}(0\;||\;TPM\_CC\_PolicyOR\;||\;digests)
  \end{aligned}$
}

\vspace{.5\baselineskip}
\noindent where

\begin{description}
  \small
  \item[$TPM\_CC\_PolicyOR$] \hfill is the policy command code
  \item[$digest_{1} \ldots digest_{n}$] \hfill is a list of valid digest values\footnotemark
  \item[$digests$] \hfill is the concatenation $digest_{1}\;||\;\ldots\;||\;digest_{n}$
\end{description}

\subsection{\texttt{TPM2\_PolicyPassword()}}

The \texttt{TPM2\_PolicyPassword()} is a deferred assertion that sets the
\texttt{policySession} $\rightarrow$ \texttt{isPasswordNeeded} property. This
indicates a caller provided \texttt{authValue} will be compared the
\texttt{authValue} of the authorized object when the policy session is used for
authorization. Once set, this property may only be unset by
\texttt{TPM2\_PolicyAuthValue() } or \texttt{TPM2\_PolicyRestart()}.

The \texttt{TPM2\_PolicyPassword} differs from \texttt{TPM2\_PolicyAuthValue} in
that the \texttt{authValue}s  are compared as a clear-text passwords, not HMACs.
In other respects, the commands are identical to the point were they actually
use the same policy command code in the \texttt{policyDigest} update:
 
\vspace{.5\baselineskip}
\noindent
\resizebox{\linewidth}{!}{%
  $policyDigest_{new} =
  \mathbf{H}(policyDigest_{old}\;||\;TPM\_CC\_PolicyAuthValue)$
}

\subsection{\texttt{TPM2\_PolicyPCR()}}

The \texttt{TPM2\_PolicyPCR()} command asserts the state of a specified set of
PCRs, and can act as a combined or deferred assertion depending on if the caller
provides and expected value for the PCR measurements, or not. If the caller
provides an expected value, the value of the specified PCR is compared
immediately to the expected value, else the validity of the PCR values will not
be known until the policy session is used for authorization. Either way, the
\texttt{policyDigest} is updated accordingly:

\vspace{.5\baselineskip}
\noindent
\resizebox{\linewidth}{!}{%
  $policyDigest_{new} =
  \mathbf{H}(policyDigest_{old}\;||\;TPM\_CC\_PolicyPCR\;||
  \;PCRs\;||\;digest)$
}

\vspace{.5\baselineskip}
\noindent where

\begin{description}
  \small
  \item[$TPM\_CC\_PolicyPCR$] \hfill is the policy command code
  \item[$pcrs$] \hfill is the bit mask of the PCR selection
  \item[$digest$] \hfill is the hash of the PCR selection values 
\end{description}

To detect changes to PCRs after the \texttt{TPM2\_PolicyPCR()} command has been
invoked, the TPM keeps track of PCR changes by incrementing a monotonically
increasing \texttt{policySession} $\rightarrow$ \texttt{pcrUpdateCounter} each
time a PCR is updated. When \texttt{TPM2\_PolicyPCR()} is invoked, and the
immediate PCR value check succeeds, the current value of the PCR update counter
is stored in the current session context. On subsequent \texttt{TPM2\_PolicyPCR}
invocations, and when the policy session is used for authorization, the value of
the PCR update counter is compared against the stored counter; the authorization
will fail unless the counter values match.


\fi

\end{document}